%% file: Hans-Dnbar.tex
\input phyzzx  \input mydef

 \date{March, 2021}
\date{March, 2021}
\titlepage
\title{On the $D_n$ Spin Vertex Models for Odd $n$}
\author{Doron Gepner}
\vskip20pt
\line{\it\hfill  Department of Particle Physics and Astrophysics, Weizmann Institute,\hfill}
 \line{\it\hfill Rehovot 76100,  Israel\hfill} 
 
 \abstract
 Solvable vertex models in two dimensions are of importance in conformal field theory, phase
 transitions and integrable models.
 We consider here the $D_n$ spin vertex models, for $n$ which is odd.
 The models involve also the anti--spinor representation. We describe here the Boltzmann weights
 for these representations using crossing symmetry from the previously known spinor representations.
 For calculation reasons we limit ourself to the $n=3$ and $n=5$ cases, which are described 
 explicitly.

 \endpage
 
Solvable vertex models in two dimensions are of interest to test such notions as universality,
conformal field theory and phase transitions. For reviews see
\REF\Baxter{R.J. Baxter, ``Exactly solved models in statistical mechanics", Dover Publications (1982).}
\REF\Wadati{M. Wadati, T. Deguchi and Y. Akutsu, Physics Reports 180 (4) (1989) 247.}
\REF\Francesco{P. Francesco, P. Mathieu and D. Senechal, ``Conformal field theory",
Springer--Verlag, Graduate texts in contemporary physics (1997).}
\REF\Ginsparg{P. Ginsparg, ``Applied conformal field theory", Les Houches lectures (1988).}
\r{\Baxter,\Wadati,\Francesco,\Ginsparg}.

Recently, the Boltzmann weights fo the SO$(k)$ spin vertex models were described
\REF\Gep{D. Gepner, Nucl.Phys. B 958 (2020) 115116.}
\REF\Dk{V. Belavin, D. Gepner and H. Wenzl, Nucl.Phys. B 959 (2020) 115160.}
\r{\Gep,\Dk}. This note is a continuation of these works. Specifically, in the case of
$D_n$ spin models, only the direct Boltzmann weights were determined. Namely,
the $S S\rarrow S S$ Boltzmann weights, where $S$ denotes the spin representation.
When $n$ is odd, there are also the anti--spinor representations, $\bar S$. Our purpose here is
to describe also the Boltzmann weights involving the spinor and anti--spinor representations.
For calculation reasons we do this, explicitly, only for $n=3$ and $n=5$.

Our solution for the Boltzmann weights is trigonometric. Note that rational solutions were known
before by a different method
\REF\Wig{E. Ogievetsky and P. Wiegmann, Phys.Lett. B 168 (1986) 360.}
\r\Wig.

We denote the simple roots of $D_n$ by $\alpha_r=\epsilon_r-\epsilon_{r+1}$, where 
$r=1,2,\ldots,n-1$ and $\alpha_n=\epsilon_{n-1}+\epsilon_n$, and $\epsilon_r$ are orthogonal unit vectors. The spin representation has the weights $\sum_{i=1}^n p_i \epsilon_i/2$ where
$p_i=\pm1$ and $\prod_{i=1}^n p_i=\pm 1$ for the spinor representation ($S$) and the anti--spinor representation ($\bar S$), respectively. We shift these weights by $1/2$ to describe the weights by
a vector $m$, where $m_i=0$ or $1$. We denote $V=S\oplus \bar S$.

A solution for the spin vertex models, which commutes with the co--product in the quantum group
$U_{q^2} (D_n)$ was described by Wenzl
\REF\Wenzl{H. Wenzl, ``Dualities for spin representation", arXiv: 2005.11299 (2020).}
\r\Wenzl. It is given by the element $C$ of End$(V\otimes V)$,
$$C^{b,c}_{m,s}=\sum_{j=1}^n \delta_{m_j,1-s_j} (-q^2)^{\{m-s\}_j} \delta_{b,\bar m_j}
\delta_{c,\bar s_j},\e$$
where 
$$\{m\}_j=\sum_{r=1}^j m_j,\e$$
and $\bar s_j$ is equal to $s_j$ except for the coordinate $j$ where it is $1-s_j$. The eigenvalues of
$C$ are
$$\lambda_j=\pm s(n-j),\qquad {\rm for\ } j=0,1,\ldots,n,\e$$
where
$$s(x)={q^{2 x}-q^{-2 x} \over q^2-q^{-2}}.\e$$

The solution $C$ maps $S\otimes S$ to $\bar S\otimes\bar  S$. Thus, we square it, where $C^2$ maps
the $S\otimes S$ to itself. We define the projection operators,
$$(P^a)^{b,c}_{m,s}=\prod_{p\neq a} \left [ {C^2-\lambda_p^2 I\over \lambda_a^2-\lambda_p^2}\right],\e$$
where $I$ is the identity map and the product is in End$(V\times V)$.

To make contact with conformal field theory we define
$$q^2=\exp\left[{\pi i\over r+2 n-2}\right],\e$$
where $r$ is the level of the $D_n$ WZW model \r\Francesco.
The solution to the Yang--Baxter equation for $S\otimes S$ was described in ref.
\r\Dk.
The Casimir of the representation $V_j$, which is the anti-symmetric product of $j$ vector representations, is given by
$$C(V_j)=C_j=j(2 n-j).\e$$
We assume that $n$ is odd. We define $h_i=n-2 i$, for $i=0,1,\ldots,(n-1)/2$. 
We define the `crossing' parameters,
$$\zeta_j=(C_{h_{j+1}}-C_{h_j})/2,\e$$
for $j=0,1,\ldots,(n-1)/2$.
We define $p(x)=q^x-q^{-x}$. The solution to the Yang Baxter equation for $S \otimes S\rarrow S \otimes S$, or $\bar S\otimes \bar S\rarrow \bar S\otimes \bar S$, is then
given by \r\Dk,
$$R_{m,s}^{a,b}(u)=\sum_{j=0}^{(n-1)/2} f_j(u) (P^{h_j})^{a,b}_{m,s},\e$$
where
$$f_a(u)=\left[ \prod_{j=1}^a p(\zeta_{j-1}-u)\right] \left[\prod_{j=a+1}^{(n-1)/2} p(\zeta_{j-1}+u)\right]
\bigg/ \left[\prod_{j=1}^{(n-1)/2} p(\zeta_{j-1})\right],\e$$
where $a=0,1,\ldots,(n-1)/2$. 

The solution eq. (9) obeys the Yang--Baxter equation (YBE),
$$\sum_{\alpha,\beta,\gamma} R_{j,k}^{\beta,\alpha}(u) R_{i,\beta}^{l,\gamma}(u+v) R_{\gamma,\alpha}^{m,s}(v)=\sum_{\alpha,\beta,\gamma} R_{i,j}^{\alpha,\beta}(v) R_{\beta,k}^{\gamma,s}(u+v) R_{\alpha,\gamma}^{l,m}(u),\e$$
provided that $i,j,k,l,m,s$ are all taken from the spinor representation or that they are all taken from the
anti--spinor representation.

We wish to describe now the general solution to the YBE, as applicable to the other representations in End($V\otimes V$). First we equate to zero,
$$R_{m,s}^{a,b}(u)=0,\e$$
where $m,a$ are in the spinor representation, and $s,b$ are in the anti--spinor representation,
or $m,a$ are in the anti--spinor representation and $s,b$ are in the spinor representation. 

To get the other cases we use the crossing symmetry,
$$R_{j,i}^{k,l}(u)=R_{\bar k,j}^{l,\bar i}(\lambda-u)\left[ {r(i) r(l)\over r(j) r(k)} \right] ,\e$$
where the crossing parameter is
$$\lambda=C_2/2,\e$$
and $r(i)$ are the crossing factors, which are as yet unknown.
Here $\bar i$ is the representation with the weight $-\lambda_i$ where $\lambda_i$ is the weight
of the representation $i$.
We also use the symmetry relation
$$R_{j,i}^{k,l}(u)=R_{k,l}^{j,i}(u).\e$$

The crossing symmetry gives us the $R$ matrix $R_{m,s}^{a,b}(u)$ where $m,b$ are in the spinor
representation and $s,a$ are in the anti--spinor representation, or vice versa. This completes
all the values of $R(u)$ in End$(V\otimes V)$. 

We still need to determine the crossing factors.
This we do on a case by case basis. The idea is to substitute substitute some unknown crossing factors, eq. (13) and then to use the YBE with various weights and to solve it for the crossing factors.
We also know the relation,
$$r(\bar i)=1/r(i).\e$$

In the case of $D_3$ we have four weights in the representation $S$ which are 
$$\lambda_1=\{1,0,0\},\ \ \lambda_2=\{0,1,0\},\ \ \lambda_3=\{0,0,1\},\ \ \lambda_4=\{1,1,1\}.\e$$ 
The weight of $\bar i$ is $\lambda_{\bar i}=\{1,1,1\}-\lambda_i$. By solving the YBE we get some values
for the crossing factors. There are several solutions. One of them is the following,
$$\{r[1],r[2],r[3],r[4]\}=\{-1,q,-q^2,1/q\}.\e$$
Substituting these values, we can see that the YBE indeed holds for all the values of the weights in
End$(V\otimes V)$.

Let us turn now to $D_5$. We find that the conservation law
$$R_{i,j}^{k,l}(u)=0\ {\rm if\ } \lambda_i+\lambda_j\neq \lambda_k+\lambda_l,\e$$
implies that the Boltzmann weights do not depend on the crossing factors, eq. (13), since
the only solution for eq. (19) for $S\bar S\rarrow \bar S S$ is when
$i$ and $j$ are equal to $k$ and $l$, or vice versa, and then the crossing factors cancel.
Thus, we can directly substitute the YBE and we find that it is, indeed, obeyed for various weights.

For $D_n$ with $n\geq 7$ our limitation of the calculation power, implies that it is not practical
to do the computation. So, we have to limit ourself to the explicit examples of $D_3$ and $D_5$.

\ack
I thank Alexander Zamolodchikov, Alexander Belavin and Paul Wiegmann for helpful discussions.

\refout

\bye

%% file: mydef.tex
\def\e{\adveq\eqno{\rm (\chapterlabel\the\equanumber)}}

\def\adveq{\global\advance\equanumber by 1}
\def\myeq{{\rm \chapterlabel.\the\equanumber}}
\def\rarrow{\rightarrow}

\def\semidirect{\mathrel{\raise0.04cm\hbox{${\scriptscriptstyle |\!}$
\hskip-0.175cm}\times}}


\def\ref#1{$^{[#1]}$}

\def\r#1{$[\rm#1]$}

\def\e{\adveq\eqno{\rm (\chapterlabel\the\equanumber)}}

\def\adveq{\global\advance\equanumber by 1}
\def\myeq{{\rm \chapterlabel.\the\equanumber}}
\def\rarrow{\rightarrow}

\def\semidirect{\mathrel{\raise0.04cm\hbox{${\scriptscriptstyle |\!}$
\hskip-0.175cm}\times}}


\def\ref#1{$^{[#1]}$}

\def\r#1{$[\rm#1]$}